\documentclass[a4paper,10pt]{article}
\usepackage[utf8]{inputenc}
\usepackage{amssymb}

\usepackage{epsfig} 
\usepackage{amsmath} 
\usepackage{amsfonts}
\usepackage{graphicx} 
\usepackage{colordvi}
\title{Universality of Short Distance Corrections to Quantum Optics}
\author{Mir Faizal$^1,^2$ and Davood Momeni$^3,^4$\\
 $^1$Department of Physics and Astronomy,\\ 
University of Lethbridge, Lethbridge, AB T1K 3M4, Canada \\
$^2$
Irving K. Barber School of Arts and Sciences, \\ University of British 
Columbia - Okanagan,\\  Kelowna, British Columbia V1V 1V7, Canada\\
$^3$ Department of Physics, College of Science, \\Sultan Qaboos University, P.O. Box 36,\\ Al-Khodh 123, Muscat, Sultanate of Oman\\
$^4$Tomsk State Pedagogical University, TSPU, 634061 Tomsk, Russia
}
\date{}
\begin{document}

\maketitle

\begin{abstract}
As quantum optical phenomena are based on Maxwell's equations, and it is becoming important 
to understand  quantum optical phenomena at short distances, so it is   important to analyze quantum optics 
using short distance corrected Maxwell's equation. 
Maxwell's action can be obtained from quantum electrodynamics using the framework of effective field theory, 
and so the leading order short distance corrections to Maxwell's action 
can also be obtained from the derivative expansion of the same
effective field theory. Such short distance corrections will be universal for all quantum optical systems, 
and they will effect all short distance quantum optical phenomena. In this paper, we will analyze the form 
of such corrections, and demonstrate the standard formalism of quantum optics can still be used 
(with suitable modifications), to analyze quantum optical phenomena from 
this short distance corrected Maxwell's actions. 
\end{abstract}

Quantum optics  is   important and   interesting both from a theoretical  and a technological 
point of view.  
Quantum optics has been used to study purely quantum mechanical properties of a system, 
such as quantum entanglement between optical states 
\cite{ab1}-\cite{1a}, and such studies  can even have technological applications for   communications 
\cite{1}-\cite{1ad}. It may be noted that it is important to understand the short distance 
behavior of such entangled  quantum systems, as the short distance corrections  of  quantum 
systems can have important  
consequences for entanglement in those quantum systems \cite{short}-\cite{short1}. 
 Laser are another interesting application of quantum optics, as a laser is  produced  
 through a process of optical 
 amplification based on the stimulated emission of electromagnetic radiation \cite{4}-\cite{4a}. 
 Laser has also been used to in the study  of biological systems and have thus found 
 applications in medicine \cite{4c}. 
Lasers can be used  to 
 construct optical switches (which can  undertake all binary logic operations),  so lasers 
may be used in  the construction of   quantum computers \cite{4b}-\cite{b4}.
 It is also possible to generate and use ultra-short    laser pulses, 
 and the short distance behavior of photons is important in understanding the behavior 
 of such ultra-short laser pulses \cite{8a}-\cite{7b}.  
 Thus, quantum optics is a vast field with many interesting applications, and it  
 is important to understand the short distance behavior of photons to better understand 
 many applications of quantum optics. 

The analysis of these and many other  quantum optical systems is done using the Maxwell's action. 
It may be noted that even though the  full theory describing the properties of light
is quantum electrodynamics,   
 the Maxwell's action can be obtained from quantum electrodynamics 
 using the framework of effective field theory \cite{1q}-\cite{q2}. 
 This is because the scale at which quantum optics operates is generally much larger than the mass of an electron, 
so we can integrate the high energy modes out of the action for quantum electrodynamics, 
and obtain the Maxwell's action as the low energy effective action for light \cite{q1}-\cite{q2}. 
However, at short distances, the Maxwell's equations will be corrected, and these corrections will become 
important for all short distance optical phenomena.   

Recently, the short distance behavior of quantum optical systems has becomes very important. 
In fact, it has been proposed that the quantum optical systems can be used to analyze quantum 
gravitational effects \cite{qg1}-\cite{qg2}.
Thus, it is important to understand  the short distance behavior 
of quantum optical systems, and this can be done by analyzing the quantum corrections to the Maxwell's action 
using the framework of  effective field theory. Thus, we can analyze the next to the leading order corrections to 
the Maxwell's action, using  the derivative expansion in the framework of effective field theories. 
The corrections to the Maxwell's action, from the derivative expansion in the framework of 
effective field theories, can be written as \cite{1q}-\cite{q2}
\begin{equation}
 S = \int d^4 x \mathcal{L} = \int d^4 x \left[
 -\frac{1}{4} F^{\mu\nu} F_{\mu\nu} + \frac{\alpha}{60 \pi m^2} F^{\mu\nu} \Box F_{\mu\nu}\right], 
\end{equation}
where $\alpha$ is the fine structure constant, $m$ is the mass of an electron, and 
$F_{\mu\nu} = \partial_\mu A_\nu - \partial_\nu A_\mu $. Now this correction will be the first 
order corrections which will become important at short distances. 
This correction will occur as an radiative correction, due to the creation of an virtual electron and 
positron pair, as the photon propagates.  
In this paper, we will analyze the effect of these corrections on the basic properties of a 
quantum optical system. Even though we will only analyze some basic properties of an optical system,
with these corrections, we would like to point out that such corrections will occur universally in all 
quantum optical systems, and will become important for all short distance effects in such quantum 
optical systems.

It may be noted that the  corrected Maxwell action given by Eq. (1) is still an free theory. 
Now we will use Eq. (1) as the fundamental action describing the corrected quantum optical system,
and observe how this changes the behavior of quantum optical phenomena. 
Thus, we can use the formalism of quantum optics to but replace Maxwell's action by this new action 
given in Eq. (1). 
The equation of motion obtained from  the corrected Maxwell's action (given by Eq. (1)), can be written as 
\begin{eqnarray}
&&\partial_{\mu}F^{\mu\nu}-\frac{\alpha}{15\pi m^2}\Box \partial_{\mu}F^{\mu\nu}=0 \label{eom}.
\end{eqnarray}
As  the Lorentz gauge for this action is 
$\partial_{\mu}F^{\mu\nu}=-\Box A^{\nu}$,  we can first   solve the equation of motion for  
  for $\xi^{\nu}=-\Box A^{\nu}$, 
and then obtain  $A^{\nu}$ as a solution for wave equation with source $\xi^{\nu}$. 
So, we observe that the equation for $\xi^{\mu}$ is a massive Klein-Gordon equation,   
$\Box \xi^{\mu}-\frac{{15\pi m^2}}{\alpha}\xi^{\mu}=0\label{eom2}. 
$
The mass terms appears as the modified action has taken into account the possibility of a photon to 
forming a virtual election and virtual positron pair, and then the virtual pair 
recombine to form the same photon. This  gives the system an effective mass,
and thus this quantum optical system is effectively described as by a massive system. 
The solution for this equation can be written as 
\begin{eqnarray}
 \xi^{\alpha}(\vec{x},t)=\int\frac{d^3k}{2\omega(\vec{k})}
\Big[\vec{\epsilon}^{\alpha}(k)b_ke^{-i\Big(\omega(k, \alpha)t-\vec{k}\cdot\vec{x}\Big)}
+\vec{\epsilon}^{\alpha^{*}}(k)b^{\dagger}_ke^{i\Big(\omega(k, \alpha)t-\vec{k}\cdot\vec{x}\Big)}
\Big], && 
\end{eqnarray}
where $\omega(k, \alpha)$ is now given by 
\begin{eqnarray}
\omega(k, \alpha)=c\sqrt{k^2-\frac{{15\pi m^2}}{\alpha}}. 
\end{eqnarray}  
Thus, the value of $\omega(k)$ changes by an $\alpha$ 
dependent factor to $\omega(k, \alpha)$. 
This is the main change that will occur due to the short distance correction 
of the Maxwell's action. However, as $\omega(k, \alpha)$ will be important  for quantum optical phenomena, 
this short distance correction will change the behavior of all those 
short distance quantum optical phenomena. It may be 
noted that for large distance quantum optical phenomena, we can again neglect the effect of these corrections, 
and obtain the standard results back. 
Furthermore,  even for this short distance corrected optical system, we can write,  $
\vec{\epsilon}^{\mu}(k)=\vec{\epsilon}^{\mu}_x(k)\hat{x}+\vec{\epsilon}^{\mu}_y(k)\hat{y},
\ \ |\vec{\epsilon}^{\mu}_x(k)|^2+|\vec{\epsilon}^{\mu}_y(k)|^2=1. $

Here we  obtain a cutoff over $k$ as $k\geq k_0\equiv  \sqrt{\frac{15\pi m^2}{\alpha}}$. 
It corresponds to the minimum of length in the system, such that 
$x_{min}\sim \sqrt{\frac{\alpha}{15\pi m^2}}$.  It had been argued that 
the existence of a minimal length provide evidence for the existence of a quantum gravitational effect 
in the optical system  \cite{qg1}-\cite{qg2}. However, we observe that even a minimal length will exist due to 
the cutoff in quantum electrodynamics, and this minimum length will occur much before the 
minimal length generated by quantum gravity. So, the detection of quantum gravitational effects 
from quantum optics needs further analysis   \cite{qg1}-\cite{qg2}. 

The solution to the equation $\Box A^{\mu}=-\xi^{\mu}$ can be written using the following 
Green's function,  
$G(\vec{x},t|\vec{x}',t')=[4\pi c^2|\vec{x}-\vec{x}'|]^{-1}\delta\Big(t-t'-(|\vec{x}-\vec{x}'|/{c})
\Big)
$
The solution for $A^{\mu}(\vec{x},t)$ can be obtained using this new Green's function. 

Now  if we consider this system is placed 
in a box with volume $V$, then we can write 
$
 A^{\mu}(\vec{x},t)=\sum_{\vec{k},s=1,2}\sqrt{\frac{2\pi \hbar c^2}{\omega(\vec{k})}}\int_{-\infty}^{\infty}
\frac{dx'^3}{|\vec{x}-\vec{x}'|}
  \Big[\vec{\epsilon}^{\mu}_{\vec k s}b_{\vec k s}e^{-i\Big(\omega(k, \alpha)\Big(t-\frac{|\vec{x}-\vec{x}'|}{c}
	\Big)-\vec{k}\cdot\vec{x}'\Big)}+H.c 
\Big] .  
$
Here $H.c$ stands for Hermitian conjugate.
 Now we observe that for the  integral given by $
 I^{\alpha}(\omega(k, \alpha),\vec{k})=\int_{-\infty}^{\infty}
{dx'^3}{|\vec{x}-\vec{x}'|^{-1}}e^{-i\vec{k}\cdot(\vec{x}-\vec{x}')+i\frac{\omega(k, \alpha)}{c}|
\vec{x}-\vec{x}'|} =\frac{2\pi}{k}\Big[\frac{\Im(\frac{\omega(k, \alpha)}{c}+k)}{\frac{\omega(k, \alpha)}{c}+k}+
\frac{\Im(\frac{\omega(k, \alpha)}{c}-k)}{\frac{\omega(k, \alpha)}{c}-k}
\Big] $. We also observe 
   that $(\omega/c)^2=k^2-k_0^2$, if 
$|k|<k_0$, and then $\omega(k, \alpha)\in\mathcal{C}$, thus  $\Im{(\omega(k, \alpha)/c\pm k)}>0$ , so  
$
|I^{\alpha}(\omega(k, \alpha),\vec{k})|={4\pi\omega(k, \alpha)}[{kck_0^2}]^{-1}
$.
So,  for this system placed in a box, we can write 
\begin{eqnarray}
 A^{\mu}(\vec{x},t)=\sum_{|\vec{k}|\geq |k_0|,s=1,2}
\frac{(2\pi)^{3/2}}{\pi k_0^2}\frac{\sqrt{\hbar\omega(k, \alpha)}}
{k}
\Big(\vec{\epsilon}^{\mu}_{\vec k s}b_{\vec k s}e^{-i(\omega(k, \alpha)t-\vec{k}\cdot\vec{x})}+H.c
\Big). && \label{Amu}
\end{eqnarray}
Now we can use this Eq. (\ref{Amu}), to obtain the corrected expression for the 
electric field $\vec E$ and magnetic field $\vec B$. 
This is the expression for the new electric and magnetic fields, obtain from Eq. (1) instead 
of the usual Maxwell's action. These new   electric and magnetic fields will depend will be corrected by 
an $\alpha$ dependent term, as the $\omega(k, \alpha)$ gets corrected by an $\alpha$ dependent term. 
This correction 
can thus be observed in all phenomena involving the  electric and magnetic fields.

Now we can write a complete set of positive energy or positive frequency solutions 
of the field $A^{\mu}(\vec{x},t)$ as
\begin{eqnarray}
&&g_{\vec k s}=\Big(\frac{(2\pi)^{3/2}}{\pi k_0^2}\frac{\sqrt{\hbar\omega(k, \alpha)}}
{k}\Big)^{-1/2}
e^{-i(\omega(k, \alpha)t-\vec{k}\cdot\vec{x})}.
\end{eqnarray}
These solutions are orthonormal,     
$
\Big(g_{\vec k s},g_{\vec k' s'}\Big)=C_k\delta(\vec k-\vec k')\delta_{ss'},\ \ 
\Big(g_{\vec k s},g^{*}_{\vec k' s'}\Big)=0.
$
The general solution to field $A^{\mu}$ is a linear combination of positive energy solutions 
$g_{\vec k s}$ and negative energy solutions $g^{*}_{\vec k' s'}$. 
Expanding the field $A^{\mu}(\vec{x},t)=A^{\mu}(x)$ in these modes, we have
\begin{eqnarray}
&&A^{\mu}(x)=\sum_{s=1,2}\int d^{3}k\Big[\vec{\epsilon}^{\mu}_{\vec k s}b_{\vec k s}
g_{\vec k s}+(\vec{\epsilon}^{\mu}_{\vec k s})^{\dagger}b^{\dagger}_{\vec k s}g_{\vec k s}^{*}
\Big].
\end{eqnarray}
Furthermore, it is possible to write 
$\vec{\epsilon}^{\mu}_{\vec k s}b_{\vec k s}=(g_{\vec k s},A^{\mu}(x))
=i\int d^3x(g_{\vec k s}^{*} \pi^{\mu}_s(x)-i\omega_k g_{\vec k s}^{*}A^{\mu}(x)). 
$
Now we consider  a complete orthonormal set 
of wave packets, such that $ 
(f_{j}^{\mu}(x),f_{j'}^{\nu}(x))=\delta^{\mu\nu}_{jj'},  \ \ ((f_{j}^{\mu}(x))^{*},f_{j'}^{\nu}(x))=0$, (with
a discrete index $j$). So, for  
$
A^{\mu}(x)=\sum_{j,s}\Big(f_{j}^{\mu}(x)b_{\vec j s}+f_{j}^{\mu *}(x)b^{\dagger}_{\vec j s}\Big) 
\label{fielddecomposition}
$, we have  $
[\vec{\epsilon}^{\mu}_{j s}b_{j s},(\vec{\epsilon}^{\nu}_{\vec j' s'})^{\star}b^{\dagger}_{\vec j'
s'}]=\delta^{\mu\nu}\delta_{ss'}\delta_{jj'},\ \ [\vec{\epsilon}^{\mu}_{j s}b_{j s},
\vec{\epsilon}^{\nu}_{j' s'}b_{j' s'}]=0.\label{commutation-j}$ 
Thus, it is possible to analyze the quantum optical phenomena form the new action given by Eq. (1), 
and use the techniques developed for the usual quantum optical  to analyze this 
new system.  

This formalism can be used to define a Fock space for this new optical system, and 
this  new system again  resembles   
a simple harmonic oscillator, in terms of annihilation and creation operators. 
So, it is possible to define quadrature operators
for this new optical system, as it done for the usual optical system.
However, as   $\omega(k)$ changes by an $\alpha$ 
dependent factor to $\omega(k, \alpha)$, the quadrature operators will also be modified accordingly. 
Thus, it would be possible to   use this short distance correction
for different optical system, and analyze the 
effect of such short distance corrections on various quantum optical phenomena. It would be interesting 
to explicitly derive an expression for coherent and squeezed states, and analyze the effect of these 
short distance corrections on the nonclassicality of quantum optical systems. As these corrections 
occur due to the corrections to the Maxwell's actions, they will effect all these quantum optical 
states, and may have some  interesting consequences for such optical systems.


\begin{thebibliography}{99}


 \bibitem{ab1} S. J. Freedman and J. F. Clauser, Phys. Rev. Lett. 28, 938  (1972)
 \bibitem{ab14}A. Aspect, P. Grangier, and G. Roger, Phys. Rev. Lett. 47, 460 (1981)

 
  \bibitem{ab16}
 G. Weihs, T. Jennewein, C. Simon, H. Weinfurter and A. Zeilinger, 
 Phys. Rev. Lett. 81, 5039 (1998)

  \bibitem{ab18} 
 A. Aspect, J. Dalibard and G. Roger,  
 Phys. Rev. Lett. 49, 1804 (1982)

\bibitem{1b}
 N. Leonhard, G. Sorelli, V. N. Shatokhin, C. Reinlein and A. Buchleitner, 
 Phys. Rev. A 97, 012321 (2018) 
\bibitem{1c}
M. H. M. Passos, W. F. Balthazar, A. Z. Khoury, M. H-Meyll, L. Davidovich and J. A. O. Huguenin, 
Phys. Rev. A 97, 022321 (2018)
\bibitem{1a}
F. A. S. Barbosa, A. S. Coelho, L. F. M. Martinez, L. O-Gutierrez, 
A. S. Villar, P. Nussenzveig and M. Martinelli,  Phys. Rev. Lett. 121, 073601 (2018)


\bibitem{1}
    L-M. Duan, M. D. Lukin, J. I. Cirac and P. Zoller, Nature 414,   413 (2001)
\bibitem{a1}
D. D. Awschalom, R.  Hanson, J. Wrachtrup and  B. B. Zhou, Nature Photonics  12, 516  (2018)  
\bibitem{b1}
    M. Bock, P. Eich, S. Kucera, M. Kreis, A. Lenhard, C. Becher and J. Eschner, 
     Nature Commun.  9,  1998 (2018) 
 \bibitem{1ad}N. Gisin and R. Thew,   Nature Photonics 1 , 165 (2007) 
\bibitem{short}D. Boyanovsky, Phys. Rev. D 97, 065008 (2018)
\bibitem{1short} D. Boyanovsky, Phys. Rev. D 92, 023527 (2015) 
\bibitem{2short}H. H. Tu and R. Orus, Phys. Rev. Lett. 107, 077204 (2011) 
\bibitem{short1}D. Nesterov and S. N. Solodukhin, JHEP 1009, 041 (2010) 
     
\bibitem{4}
T. H. Maiman, Nature 187,   493 (1960)
\bibitem{4ab}M. Bertolotti, Masers and Lasers: An Historical Approach, Hilger, Bristol (1983)
\bibitem{4ab1} A. L. Schawlow and  C. H. Townes,  Phys. Rev. 112, 1940 (1958)
\bibitem{4ab2}A. J. Kastler,   Phys. Radium, Paris 11, 255 (1950)
\bibitem{4a}
D. A. G. Deacon, L. R. Elias, J. M. J. Madey, G. J. Ramian, H. A. Schwettman and T. I. Smith
Phys. Rev. Lett. 38, 892 (1977)
\bibitem{4c}     V. S. Letokhov, Nature 316,   325 (1985)


\bibitem{4b}    S. D. Smith, Nature   316, 324 (1985) 
\bibitem{b4}   T. D. Ladd, F. Jelezko, 
R. Laflamme, Y. Nakamura, C. Monroe and J. L. O'Brien,
Nature   464,   45  (2010)

\bibitem{8a}S. Kaufmann, D. Schwarzer, C. Reichardt, A. M. Wodtke and O. Bunermann, Nature Commun 5,  5373 (2014)
\bibitem{8b}R Bohme, S Pissadakis, M Ehrhardt, D Ruthe and K Zimmer,  J. Phys. D, Appl. Phys. 39, 1398 (2006)
\bibitem{7a}D. Del Sorbo, D. Seipt, T. G. Blackburn, A. G. R. Thomas, C. D. Murphy, J. G. Kirk and 
C. P. Ridgers, Phys. Rev. A 96, 043407 (2017) 
\bibitem{7b}V. Y. Kharin, D. Seipt, and S. G. Rykovanov
Phys. Rev. Lett. 120, 044802 (2018)


     
\bibitem{1q}C.P. Burgess,  Ann. Rev. Nucl. Part. Sci. 57, 329 (2007)
\bibitem{2q} Z.~Fodor, C.~Hoelbling, S.~D.~Katz, L.~Lellouch, A.~Portelli, K.~K.~Szabo and B.~C.~Toth,
  Phys. Lett. B 755, 245 (2016)
\bibitem{q1} X. Kong and F. Ravndal,  Phys. Rev. Lett. 79, 545 (1997)
\bibitem{q2}C. Itzykson and J B. Zuber, Quantum Field Theory, McGraw-Hill, New York (1980)
\bibitem{qg1}
 I. Pikovski, M. R. Vanner, M. Aspelmeyer, M. Kim and C. Brukner, Nature Physics 8, 393-397 (2012) 
 \bibitem{qg4}
   S.~Dey, A.~Bhat, D.~Momeni, M.~Faizal, A.~F.~Ali, T.~K.~Dey and A.~Rehman,
  Nucl. Phys. B { 924}, 578 (2017)
  \bibitem{qg6}
    P.~Bosso, S.~Das, I.~Pikovski and M.~R.~Vanner,  Phys. Rev. A {  96},  023849 (2017)
  \bibitem{qg12}
    C.~C.~Gan, C.~M.~Savage and S.~Z.~Scully,  Phys. Rev. D { 93}, no. 12, 124049 (2016)
 \bibitem{qg8}
  M. Zych, F. Costa, I. Pikovski and C. Brukner, Nature Phys. 8, 393  (2012)  
\bibitem{qg2}
  M.~Khodadi, K.~Nozari, S.~Dey, A.~Bhat and M.~Faizal,  Sci. Rep. 8,  1659 (2018)


\end{thebibliography}
\end{document}